\documentclass[a4paper,pre,aps,showkeys,floatfix,showpacs,twocolumn,%
superscriptaddress,amsmath,amssymb]{revtex4}

\usepackage{dsfont,graphicx}
\usepackage{bm}
\usepackage{comment}
\usepackage{mathptmx}
\usepackage{upgreek}
\usepackage{microtype}

\newcommand{\longitudinal}{l}
\newcommand{\transverse}{$\perp$}
\newcommand{\interference}{interf}
\newcommand{\grad}{\boldsymbol{\nabla}}
\newcommand{\bfA}{{\mathbf A}}
\newcommand{\bfB}{{\mathbf B}}
\newcommand{\bfBf}{{\mathbf B}_{\text{far}}}
\newcommand{\bfBn}{{\mathbf B}_{\text{near}}}
\newcommand{\bfE}{{\mathbf E}}
\newcommand{\bfEf}{{\mathbf E}_{\text{far}}}
\newcommand{\bfEi}{{\mathbf E}_{\text{inter}}}
\newcommand{\bfEn}{{\mathbf E}_{\text{near}}}
\newcommand{\bfI}{{\mathbf I}}
\newcommand{\bfJ}{{\mathbf J}}
\newcommand{\bfL}{{\mathbf L}}
\newcommand{\Lz}{L_z}
\newcommand{\Lxi}{L_x^{\text{\interference}}}
\newcommand{\Lyi}{L_y^{\text{\interference}}}

\newcommand{\Lzl}{L_z^{\text{\longitudinal}}}
\newcommand{\Lzt}{L_z^{\text{\transverse}}}
\newcommand{\bfP}{{\mathbf P}}

\newcommand{\bfPl}{{\mathbf P}^{\text{\longitudinal}}}
\newcommand{\bfPt}{{\mathbf P}^{\text{\transverse}}}
\newcommand{\bfS}{{\mathbf S}}
\newcommand{\Sz}{S_z}
\newcommand{\Sxi}{S_x^{\text{\interference}}}
\newcommand{\Syi}{S_y^{\text{\interference}}}

\newcommand{\Szl}{S_z^{\text{\longitudinal}}}
\newcommand{\Szt}{S_z^{\text{\transverse}}}

\newcommand{\Wl}{W^{\text{\longitudinal}}}
\newcommand{\Wt}{W^{\text{\transverse}}}

\newcommand{\bfj}{{\mathbf j}}
\newcommand{\bfjf}{{\mathbf j}_{\text{far}}}

\newcommand{\bfjl}{{\mathbf j}_{\text{far}}^{\text{\longitudinal}}}

\newcommand{\bflf}{{\mathbf l}_{\text{far}}}
\newcommand{\bfli}{{\mathbf l}_{\text{far}}^{\text{\interference}}}
\newcommand{\bfll}{{\mathbf l}_{\text{far}}^{\text{\longitudinal}}}
\newcommand{\bflt}{{\mathbf l}_{\text{far}}^{\text{\transverse}}}
\newcommand{\bfp}{{\mathbf p}}
\newcommand{\bfpf}{{\mathbf p}_{\text{far}}}
\newcommand{\bfpi}{{\mathbf p}_{\text{far}}^{\text{\interference}}}
\newcommand{\bfpl}{{\mathbf p}_{\text{far}}^{\text{\longitudinal}}}
\newcommand{\bfpt}{{\mathbf p}_{\text{far}}^{\text{\transverse}}}
\newcommand{\bfpn}{{\mathbf p}_{\text{next}}}
\newcommand{\bfpr}{{\mathbf p}_{\text{rest}}}
\newcommand{\bfr}{{\mathbf r}}
\newcommand{\bfsf}{{\mathbf s}_{\text{far}}}
\newcommand{\bfsi}{{\mathbf s}_{\text{far}}^{\text{\interference}}}
\newcommand{\bfsl}{{\mathbf s}_{\text{far}}^{\text{\longitudinal}}}
\newcommand{\bfst}{{\mathbf s}_{\text{far}}^{\text{\transverse}}}
\newcommand{\w}{u}
\newcommand{\wf}{\w_{\text{far}}}
\newcommand{\wi}{\w_{\text{far}}^{\text{\interference}}}
\newcommand{\wl}{\w_{\text{far}}^{\text{\longitudinal}}}
\newcommand{\wt}{\w_{\text{far}}^{\text{\transverse}}}
\newcommand{\f}{f}
\newcommand{\F}{F}
\newcommand{\bfer}{\hat{\mathbf r}}
\newcommand{\bfetheta}{\hat{\pmb{\uptheta}}}
\newcommand{\bfephi}{\hat{\pmb{\upphi}}}

\newcommand{\calA}{{\cal A}}
\newcommand{\calC}{{\cal C}}

\newcommand{\Res}[1]{\operatorname{\underset{\text{$#1$}}{Res}}}

\begin{document}

\title{Mechanical properties of the radio frequency field
  emitted by an antenna array}

\author{H.\,Then}
 \email{holger.then@uni-oldenburg.de}
 \affiliation{%
 Institute of Physics, 
 Carl von Ossietzky Universit\"at Oldenburg,
 D-261\,11 Oldenburg,
 Germany}%

\author{B.\,Thid\'e}
 \altaffiliation[Also at ]{LOIS Space Centre, V\"axj\"o University,
 SE-351\,95 V\"axj\"o, Sweden}
 \email{bt@irfu.se}
 \affiliation{%
 Swedish Institute of Space Physics,
 \AA ngstr\"om Laboratory,
 P.\,O.\,Box 537,
 SE-751\,21 Uppsala,
 Sweden}%

\begin{abstract}

Angular momentum densities of electromagnetic beams are connected to
helicity (circular polarization) and topological charge (azimuthal
phase shift and vorticity). Computing the electromagnetic fields
emitted by a circular antenna array, analytic expressions are found
for the densities of energy, linear and angular momentum in terms of
helicity and vorticity. It is found that the angular
momentum density can be separated into spin and orbital
parts, a result that is known to be true in a beam
geometry. The results are of importance for information-rich
radio astronomy and space physics as well as novel radio, radar,
and wireless communication concepts.

\end{abstract}


\maketitle

\section{Introduction}

As Maxwell showed, any interaction between electromagnetic (em)
radiation and matter is inevitably accompanied by an exchange of
momentum.  In addition to linear momentum, Poynting
\cite{Poynting:PRSL:1909} inferred from a mechanical analogy that
circularly polarized light must carry spin angular momentum and Allen et
al.\ \cite{Allen&al:PRA:1992} argued that Laguerre-Gaussian laser modes
must carry orbital angular momentum.  The exchange of spin angular
momentum between radiation and a mechanical body was demonstrated
experimentally by Beth in 1935 \cite{Beth:PR:1936}, while the 
transfer of em orbital angular momentum to mechanical orbital
angular momentum of particles was experimentally demonstrated only
recently
\cite{Garces-Chavez&al:JOA:2004,Babiker&al:PRL:1994,Andersen:PRL:2006}.

During the past decade the application of em spin and orbital angular
momentum has come to the fore in optics \cite{Franke-Arnold&al:LPR:2008}
and in atomic and molecular physics \cite{Cohen-Tannoudji:RMP:1998}.
The orbital angular momentum of electromagnetic beams now takes an
important position in various fields of research and applications,
ranging from manipulating and orienting small particles
\cite{Ladavac&Grier:OE:2004}, astronomy and astrophysics
\cite{Harwit:APJ:2003,Swartzlander&al:OE:2008,Anzolin&al:AA:2008,Elias:AA:2008},
novel wireless communication concepts
\cite{Gibson&al:OE:2004,Lin&al:AO:2007}, quantum entanglement
\cite{Mair&al:N:2001}, cryptography, and quantum computation
\cite{Vaziri&al:JOB:2002}.  Applications for the use of the orbital
angular momentum in the radio domain, allowing digital manipulation
under software control, were proposed in
Ref.~\onlinecite{Thide&al:PRL:2007}, including radio astronomy and space
physics
\cite{Stal&al:PRL:2007,Panda&al:JCAP:2007,Istomin:PLA:2002,Paterson:PRL:2005,Thide:MMWP:2004,Thide&al:PRL:2005,Khotyaintsev&al:SP:2006,Thide:PPCF:2007,Norin&al:PRL:2009,Leyser&al:PRL:2009,Mendonca&al:PRL:2009}.

In the literature, different views can be found concerning the
possibility of separating the spin and orbital angular momentum; see
Refs.~\onlinecite{Gottfried:Book:1966,Jauch&Rohrlich:Inbook:1980,Cohen-Tannoudji&al:Inbook:1989,Barnett:JOB:2002}.
In computing the electromagnetic beams emitted by a circular antenna
array, and expressing the angular momentum density explicitly in terms
of helicity and vorticity, the present paper shows that the angular
momentum of the emitted beams can be clearly separated into parts that
belong to helicity and vorticity, respectively.  The separation is
associated with the separation of angular momentum into spin and orbital
parts, a result that has been previously shown for optical beams
\cite{Allen&al:PO:1999,Barnett:JOB:2002}.  Crucial for the analytic
computation is the fact that the radiated densities of energy, linear
momentum, and angular momentum are homogeneous functions of distance.
This allows to neglect certain terms in the expressions for these
quantities.  Since helicity and vorticity are uniquely linked to the
phases of the transverse field components, it is possible to emit and
detect em beams with spin and orbital angular momentum via transverse
field triangulation, dealing solely with first order quantities which
can easily be utilized without the need to measure weak longitudinal
components.

In order to simplify the present analysis, an approximation is made in
the geometrical setup by assuming an array with a continuum of
infinitely many elemental antennas \cite{Josefsson&Persson:Book:2006}.

\section{Setup} \label{sec:A}

The vector potential of a single or crossed electric dipole antenna
that is in harmonic oscillation with frequency $\omega$ can be
expressed as
\begin{align}
  \bfA(\bfr,t)=-\frac{\mu_0d}{8\pi}\bfJ(t)\frac{e^{ikr}}{r}\,,
\end{align}
cf.\ Eqs.~(9.16) and (9.27) in \cite{Jackson:Book:1998}.  Here $i$ is the
imaginary unit, $k$ is the magnitude of the wave vector, $\bfer$ is the
unit vector in the direction of $\bfr$, $\mu_0$ is the magnetic
permeability, $d$ is the length of each dipole antenna, and $\bfJ$ is
the antenna current.

Taking the potentials and fields to be complex valued, the harmonic time
dependence is attributed by multiplying a stationary antenna current
$\bfI$ with the appropriate complex factor,
\begin{align}
  \bfJ(t)=\bfI e^{-i\omega t}\,.
\end{align}
The physical fields result from taking the real parts.

\begin{figure}[t]
  \begin{minipage}{\columnwidth}
    \includegraphics[width=.49\columnwidth]{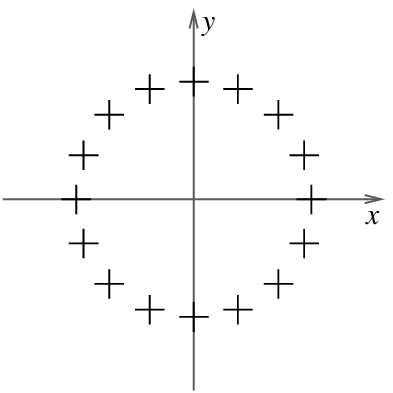}
    \hfill
    \includegraphics[width=.49\columnwidth]{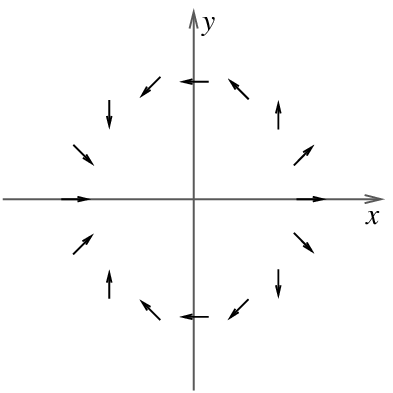}
  \end{minipage}
  \caption{%
 Plots of the $N$ crossed dipole antennas and their currents.
 The left panel displays the positions and orientations of the
 crossed dipole antennas, and the right panel displays the
 current vector at each crossed dipole for $N=16$, $h=1$, $l=-2$,
 and $t=0$.}
\label{fig:1}
\end{figure}%

Placing $N$ crossed dipole antennas on a circle at the positions
\begin{align}
  \bfr_n=\begin{pmatrix}R\cos(2\pi n/N)\\
   R\sin(2\pi n/N)\\0
  \end{pmatrix}, \quad n=1\ldots N\,,
\end{align}
where $R$ is the radius of the antenna array [see Fig.~\ref{fig:1}
(left)], and feeding the $n$th antenna element with the current
\begin{align}
  \bfI_n=\frac{1}{\sqrt{2}}\begin{pmatrix}I\\ihI\\0
  \end{pmatrix}e^{2\pi iln/N}, \quad n=1\ldots N\,,
\end{align}
[see Fig. \ref{fig:1} (right)], the total vector potential reads
\begin{align}
  \bfA=\frac{1}{\sqrt{2}}\begin{pmatrix}1\\ih\\0
  \end{pmatrix}\calA(r)\calC(\theta,\phi)
  \label{A}
\end{align}
with
\begin{align}
  \calA=-\frac{I\mu_0d}{8\pi}\frac{e^{ikr}}{r}
\end{align}
and
\begin{align}
  \calC=\sum_{n=1}^N e^{2\pi iln/N}e^{-ik\bfer\cdot\bfr_n}\,,
\end{align}
where $I$ is the amplitude of the antenna currents, $h=\pm1$ is the helicity
associated with circular polarization, and $l\in\mathds{Z}$ is the
topological charge. The topological charge attributes an azimuthal
phase shift of $\exp(2\pi iln/N)$ to the antenna currents and is
associated with vorticity.

If the number of antennas $N$ is sufficiently large (typically ${N\geq16}$),
the sum in $\calC$ can be approximated by a Riemann integral and expressed
by its residue in the complex $\zeta$ plane,
\begin{align}
  \calC(\theta,\phi)&\simeq\frac{N}{2\pi}\int_{0}^{2\pi}d\Phi
  e^{il\Phi}e^{-ikR\sin\theta\cos(\Phi-\phi)} \label{C} \\
  &=e^{il\phi}N\Res{\zeta=0}\frac{1}{l}
  e^{-ikR\sin\theta\cosh(\frac{1}{l}\ln\zeta)}\,, \label{C2}
\end{align}
yielding
\begin{align}
  \frac{\partial}{\partial\phi}\ln\calC=il \label{dCdphi}
\end{align}
and
\begin{align}
  \frac{\partial}{\partial\theta}\ln\calC=\text{real} \label{dCdtheta}
\end{align}
which implies that $\calC^*\calC$ is independent of $\phi$.

\section{Fields}

From the vector potential follows the magnetic field
\begin{align}
  \bfB=\grad\times\bfA\,.
\end{align}
Separating the magnetic field into its far and
its near field terms,
\begin{align}
  \bfB=\bfBf+\bfBn\,,
\end{align}
results in
\begin{align}
  \bfBf=ik\bfer\times\mathbf A
\end{align}
and
\begin{align}
  \bfBn=e^{ikr}\grad\times({\bfA}e^{-ikr})\,,
\end{align}
where $e^{-ikr}\bfBf$ and $e^{-ikr}\bfBn$ are homogeneous functions in
$r$ of degree $-1$ and $-2$, respectively.

The electric field outside the sources is given by
\begin{align}
  \bfE=\frac{ic}{k}\grad\times\bfB\,,
\end{align}
where $c$ is the speed of light.
The electric field results in
\begin{align}
  \bfE=\bfEf+\bfEi+\bfEn
\end{align}
with
\begin{align}
  \bfEf=ikc[\bfA-(\bfer\cdot\bfA)\bfer]\,,
\end{align}
\begin{align}
  \bfEi=-c[\bfer(\grad\cdot(\bfA e^{-ikr})) +\grad(\bfer\cdot\bfA
  e^{-ikr})]e^{ikr}\,,
\end{align}
and
\begin{align}
 \bfEn=\frac{ic}{k}[\grad(\grad\cdot({\bfA}e^{-ikr}))
 -\grad^2({\bfA}e^{-ikr})]e^{ikr}\,.
\end{align}
The individual terms of the electric field are called the far,
$\bfEf$, the intermediate, $\bfEi$, and the near electric field,
$\bfEn$. Apart from the complex phase, they are homogeneous functions
in $r$ of degree $-1$, $-2$, and $-3$, respectively.

\section{Densities of energy and momenta} \label{sec:w,p,j}

Imagine a set of concentric spheres centered on the radiation source at
the origin.  If a physical quantity $Q$ is conserved, the total flow of
$Q$ through the sphere surfaces must be the same for each sphere.  This
is only possible if the $r^2$ dependence of the surface area cancels
with an $r^{-2}$ dependence of the density of the conserved quantity.
Hence, the density of any conserved quantity $Q$ that propagates
radially with a constant speed is required to be homogeneous in $r$ of
degree $-2$.

The em field energy density is defined by
\begin{align}
  \w=\frac{\epsilon_0}{4}(\bfE\cdot\bfE^*+c^2\bfB\cdot\bfB^*)\,,
\end{align}
where $\epsilon_0$ is the electric permittivity of free space.  The
conservation law of energy imposes the condition that the radiated
energy density must be homogeneous in $r$ of degree $-2$.  This
condition is fulfilled by the leading order term of the energy
density,
\begin{align}
  \wf=\frac{\epsilon_0}{4}(\bfEf\cdot\bfEf^*+c^2\bfBf\cdot\bfBf^*)\,.
\end{align}

The linear momentum density is given by the Poynting vector
\begin{align}
  \bfp=\frac{\epsilon_0}{2}\Re[\bfE\times\bfB^*]\,,
\end{align}
which can be split
into homogeneous functions in $r$ of degree $-2$, $-3$, and higher
order terms,
\begin{align}
  \bfp=\bfpf+\bfpn+\bfpr
\end{align}
with
\begin{align}
  \bfpf=\frac{\epsilon_0}{2}\Re[\bfEf\times\bfBf^*]\,,
\end{align}
\begin{align}
  \bfpn=\frac{\epsilon_0}{2}\Re[\bfEf\times\bfBn^*+\bfEi\times\bfBf^*]
\end{align}
and
\begin{align}
  \bfpr=O(1/r^4)\,.
\end{align}
Due to the conservation of linear momentum,
the radiated linear momentum density is given by the far field terms,
$\bfEf$ and $\bfBf$, only.

The angular momentum density is defined by
\begin{align}
  \bfj=\bfr\times\bfp\,.
\end{align}
Because of the conservation law of angular momentum the radiated
angular momentum density is homogeneous in $r$ of degree $-2$. This
has important consequences. Namely, the pure far field does not
contribute to the radiated angular momentum density, since
$\bfr\times\bfpf$ would then be a homogeneous function in $r$ of degree
$-1$, but this term vanishes due to the fact that $\bfpf$ is directed
along $\bfr$. Instead, it is the next order term,
\begin{align}
  \bfjf=\bfr\times\bfpn\,,
\end{align}
that yields the radiated angular momentum density.

Plugging the vector potential expression (\ref{A}) into the expressions
for the magnetic and electric fields and evaluating the radiated
densities of energy, linear momentum, and angular momentum results in
\begin{align}
  \wf=\frac{k^2}{4\mu_0} \Big(\frac{I\mu_0d}{8\pi
    r}\Big)^2(1+\cos^2\theta)(\calC^*\calC)\,,
  \label{w}
\end{align}
\begin{align}
  \bfpf=\frac{\bfer k^2}{4c\mu_0} \Big(\frac{I\mu_0d}{8\pi
    r}\Big)^2(1+\cos^2\theta)(\calC^*\calC)\,,
\end{align}
and
\begin{multline}
  \bfjf=\frac{-\bfetheta k}{4c\mu_0}\Big(\frac{I\mu_0d}{8\pi r}\Big)^2
  \Big\{ h\Big( 2\sin\theta(\calC^*\calC)
  -\cos\theta\frac{\partial}{\partial\theta}(\calC^*\calC) \Big)
  \\
  +l\frac{1+\cos^2\theta}{\sin\theta}(\calC^*\calC) \Big\}\,,
  \label{j}
\end{multline}
where $\bfer$, $\bfetheta$, and $\bfephi$ are the unit vectors of
spherical coordinates. Since the density of angular momentum scales
with distance like $r^{-2}$, the angular momentum is transported all
the way out to infinity, a fact that is not always well appreciated.

\section{Spin and orbital angular momentum} \label{sec:s,l}

Separating the angular momentum density in terms that depend on the
topological charge $l$ only and in terms that depend on the helicity
$h$,
\begin{align}
  \bfjf=\bflf+\bfsf\,,
\end{align}
allows to identify the topological charge
(vorticity) with the orbital angular momentum density
\begin{align}
  \bflf=l\frac{-\bfetheta k^2}{4\omega\mu_0}\Big(\frac{I\mu_0d}{8\pi
    r}\Big)^2
  \Big(\frac{1+\cos^2\theta}{\sin\theta}\Big)(\calC^*\calC)\,,
\end{align}
and the helicity (polarization) with the spin orbital angular
momentum density
\begin{align}
  \bfsf=h\frac{-\bfetheta k^2}{4\omega\mu_0}\Big(\frac{I\mu_0d}{8\pi
    r}\Big)^2 \Big( 2\sin\theta(\calC^*\calC)
  -\cos\theta\frac{\partial}{\partial\theta}(\calC^*\calC) \Big)\,.
  \label{s}
\end{align}

\begin{figure}[t]
  \begin{minipage}{\columnwidth}
    \includegraphics[width=.49\columnwidth]{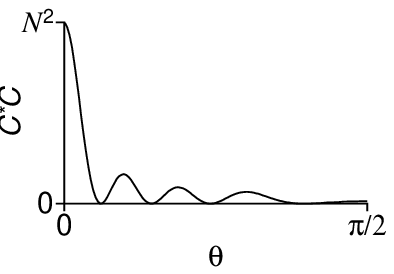}
    \hfill
    \includegraphics[width=.49\columnwidth]{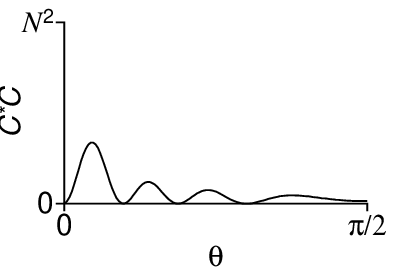}
  \end{minipage}
  \caption{%
 Plots of the function $\calC^*\calC$. Notice the influence of
 $l$ on the graph. Here $l=0$ (left) and $l=\pm1$ (right).}
\label{fig:2}
\end{figure}%

Surprisingly, the orbital angular momentum density is easier to
describe than the density of the spin angular momentum. The density of
the orbital angular momentum is completely independent of the
helicity, but the spin angular momentum density depends via
$\calC^*\calC$ on the topological charge. If the topological charge
vanishes, $\calC^*\calC$ has its maximum at $\theta=0$, while for
any non-vanishing topological charge, $\calC^*\calC$ is zero at
$\theta=0$,
\begin{align}
  (\calC^*\calC)_{|\theta=0} =\begin{cases}N^2&\text{if
      $l=0$\,,}\\0&\text{if $l\not=0$\,,}\end{cases}
\end{align}
and
\begin{align}
  \frac{\partial}{\partial\theta}(\calC^*\calC)_{|\theta=0}=0\,.
\end{align}
Figure \ref{fig:2} shows graphs of $\calC^*\calC$. For all plots, an array
radius of $R=2\lambda$ was used, where $\lambda={2\pi}/{k}$ is the
wavelength.

\begin{figure}[t]
  \begin{minipage}{\columnwidth}
    \includegraphics[width=.32\columnwidth]{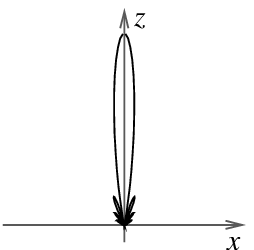}
    \hfill
    \includegraphics[width=.32\columnwidth]{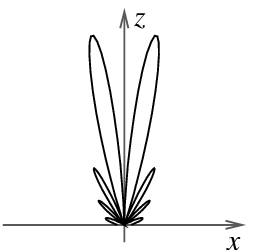}
    \hfill
    \includegraphics[width=.32\columnwidth]{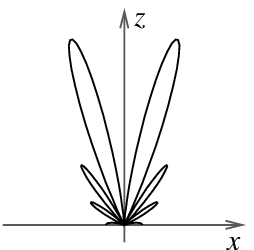}
  \end{minipage}
  \caption{%
 Radiation patterns of the linear momentum for radio beams
 generated by a circular antenna array with radius $2\lambda$;
 all antennas are directly over ground. Without ground the
 electromagnetic waves would also be emitted symmetrically into the
 lower half space.
 The patterns are for $l=0$ (left), $l=\pm1$ (middle),
 and $l=\pm2$ (right).}
\label{fig:3}
\end{figure}%

\section{Radiation patterns}

Radiation patterns of the energy (linear momentum) are shown in
 Fig.~\ref{fig:3}.
If the topological charge vanishes, i.e.\ $l=0$, the particular antenna array
chosen here emits a beam that has maximum intensity on the $z$ axis. If the
topological charge is non-vanishing, $l\not=0$, the fields of the beam
cancel on the $z$ axis, resulting in a doughnut shaped beam profile
that becomes wider for larger absolute values of the topological
charge.

The far-field approximation of the Poynting vector is proportional to
the energy density and points from the antenna array radially outward,
\begin{align}
  \bfpf=\frac{\bfer}{c}\wf\,.
\end{align}
Therefore, the radiation pattern of the linear momentum is identical
to that of the energy.

\begin{figure}[t]
  \begin{minipage}{\columnwidth}
    \includegraphics[width=.32\columnwidth]{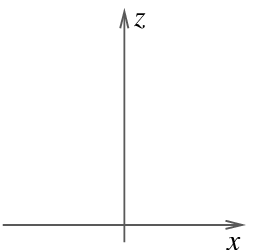}
    \hfill
    \includegraphics[width=.32\columnwidth]{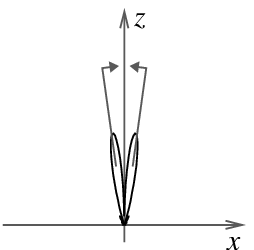}
    \hfill
    \includegraphics[width=.32\columnwidth]{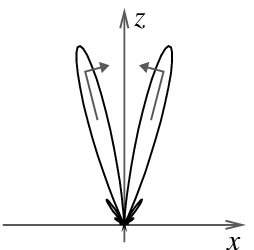}
  \end{minipage}
  \caption{%
 Radiation patterns for the orbital angular momentum.
 The direction of the orbital angular momentum density vectors
 is in the direction of $-\bfetheta l$ as displayed by the
 arrows $\Rsh$ and $\Lsh$. The patterns are for $l=0$ (left),
 $l=1$ (middle), and $l=2$ (right).}
\label{fig:4}
\end{figure}

\begin{figure}[t]
  \begin{minipage}{\columnwidth}
    \includegraphics[width=.32\columnwidth]{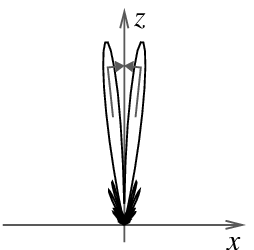}
    \hfill
    \includegraphics[width=.32\columnwidth]{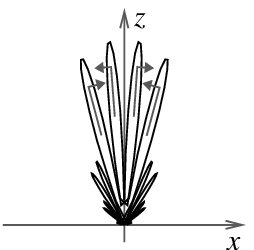}
    \hfill
    \includegraphics[width=.32\columnwidth]{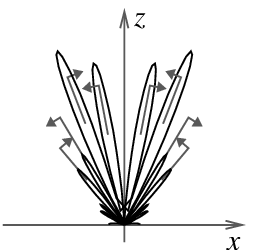}
  \end{minipage}
  \caption{%
 Radiation patterns for the spin angular momentum.
 The directions of the spin angular momentum density vectors
 alternate between $-\bfetheta h$ and $+\bfetheta h$ as
 displayed by the arrows $\Rsh$ and $\Lsh$.
 The patterns are for positive helicity, $h=+1$, and
 topological charge $l=0$ (left),
 $l=\pm1$ (middle), and $l=\pm2$ (right).}
\label{fig:5}
\end{figure}

Radiation patterns of the orbital and spin angular momentum are
displayed in Figs.~\ref{fig:4} and \ref{fig:5}, respectively.
The orbital angular momentum density always points in the direction of
$-\bfetheta l$, whereas the spin angular momentum density changes its sign
from cone to cone. Hence, the directions of the spin angular momentum
density vectors alternate between $-\bfetheta h$ and $+\bfetheta h$.

Often, spin is identified with polarization. We emphasize that this is
not allowed for the spin angular momentum {\em density}; only the
total spin angular momentum can be identified with polarization. For
instance, taking $l=0$ and $h=1$, the spin angular momentum density
vanishes on the $z$ axis, see Fig. \ref{fig:5} (left), whereas the
electric and magnetic fields are maximal on the $z$ axis and are
right-hand circular polarized.

\section{Fluxes}

The total fluxes of energy, linear momentum, orbital, and spin angular
momentum follow from integrating the densities. Let $S$ be a closed
surface with the antenna array inside it. The total fluxes through
the surface are
\begin{align}
  \frac{d\F}{dt}=\int_S \f cr^2d\Omega\,,
\end{align}
where $\F$ stands for the total flux $W$, $\bfP$, $\bfL$, or $\bfS$ and
$\f$ for the corresponding density $\wf$, $\bfpf$, $\bflf$, or $\bfsf$,
respectively.
Because of the rotational symmetry around the $z$ axis, the orbital
and spin angular momenta, $\bfL$ and $\bfS$, are both parallel to the $z$
axis.  The linear momentum is radiated symmetrically in the positive
and negative $z$ directions. Thus, the net vanishes,
\begin{align}
  \bfP=\mathbf{0}\,,
\end{align}
and the angular momentum of our electromagnetic beams becomes an
intrinsic but non-local property.

Expressing the orbital and spin angular momentum in multiples of the
total energy, we obtain \cite{Allen&al:PRA:1992,Barnett:JOB:2002}
\begin{align}
  \frac{L_z}{W}=\frac{dL_z/dt}{dW/dt}=\frac{l}{\omega}
\end{align}
and
\begin{align}
  \frac{S_z}{W}=\frac{dS_z/dt}{dW/dt}=\frac{h}{\omega}\,.
\end{align}

\section{Detecting the torque}

\begin{figure}[t]
  \begin{minipage}{\columnwidth}
    \includegraphics[width=\columnwidth]{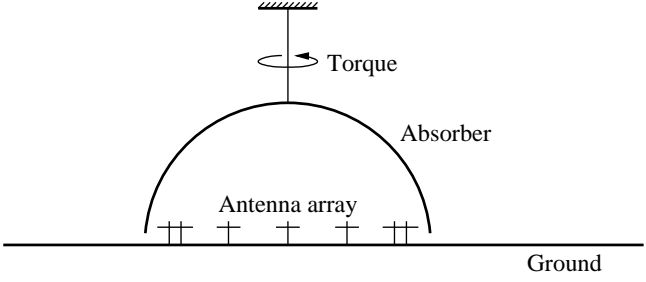}
  \end{minipage}
  \caption{%
 A radio experiment that measures the torque of the
 total angular momentum flux.}
\label{fig:6}
\end{figure}

The torque of the angular momentum flux can, at least in principle, be
directly measured with an absorber that surrounds the antenna array in
the upper half space; see Fig.~\ref{fig:6}.
If the antenna array is near ground with reflectivity
$\rho_{\text{g}}={I_{\text{refl}}}/{I_{\text{in}}}$, and if the
absorber has a reflectivity of $\rho_{\text{a}}$ and a transmittance
of $T_{\text{a}}={I_{\text{trans}}}/{I_{\text{in}}}$, the torque
is
\begin{align}
 \frac{d(L_z+S_z)}{dt}=\alpha\frac{l+h}{\omega}\frac{dW}{dt}\,,
\end{align}
where ${dW}/{dt}$ is the power that is radiated by the antenna
array and $\alpha$ is determined by the reflectivities and the
transmittance
\begin{align}
  \alpha=\frac{1+\rho_{\text{g}}}{2}(1-T_{\text{a}}-\rho_{\text{a}})
  \frac{1}{1-\rho_{\text{a}}}\,.
\end{align}

\section{Radio waves associated with longitudinal currents}
\label{sec:long}

Based on an idea put forward by Carozzi \cite{Carozzi:LOIS:2008}, let us
replace the transverse crossed dipoles of the circular antenna array by
electric dipoles that point in the longitudinal direction and feed the
longitudinal dipole antennas with the currents
\begin{align}
  \bfI_n = \frac{1}{\sqrt{2}} \begin{pmatrix} 0 \\ 0 \\ 1
  \end{pmatrix} e^{2\pi il\frac{n}{N}}, \quad n=1\ldots N\,.
\end{align}
Then the vector potential reads
\begin{align}
  \bfA = \frac{1}{\sqrt{2}} \begin{pmatrix} 0 \\ 0 \\ 1 \end{pmatrix}
  \calA(r)\calC(\theta,\phi)\,,
\end{align}
where $\calA$ and $\calC$ are as given in Section~\ref{sec:A}.

Evaluating the densities of energy, linear momentum, and angular
momentum, we obtain
\begin{align}
  \wl = \frac{k^2}{4\mu_0} \Big(\frac{I\mu_0d}{8\pi r}\Big)^2
  (1-\cos^2\theta) (\calC^*\calC)\,,
\end{align}
\begin{align}
  \bfpl = \frac{\bfer k^2}{4c\mu_0} \Big(\frac{I\mu_0d}{8\pi r}\Big)^2
  (1-\cos^2\theta) (\calC^*\calC)\,,
\end{align}
and
\begin{align}
  \bfjl = l \frac{-\bfetheta k^2}{4\omega\mu_0}
  \Big(\frac{I\mu_0d}{8\pi r}\Big)^2
  \Big(\frac{1-\cos^2\theta}{\sin\theta}\Big) (\calC^*\calC)\,.
\end{align}

It is useful to compare these densities with those given in Section
\ref{sec:w,p,j}, Eqs. (\ref{w})--(\ref{j}). The essential difference
is in the angular momentum density. For pure longitudinal currents,
there is only an orbital part,
\begin{align}
  \bfjl=\bfll\,,
\end{align}
and the spin part vanishes,
\begin{align}
  \bfsl=\mathbf{0}\,.
\end{align}
This shows that electromagnetic waves associated
with pure longitudinal currents can carry orbital, but no spin angular
momentum; cf.\ the discussion of longitudinal plasma modes in
Ref.~\onlinecite{Mendonca&al:PRL:2009}.

Because of the term $(1-\cos^2\theta)$ the beams of electromagnetic
waves associated with longitudinal currents have a deep null along
their axis of propagation which gives rise to a doughnut shaped beam
profile, even in the absence of any topological charge.

The LOIS radio facility \cite{Rothkaehl&al:JASTP:2008,LOIS}, currently
under construction in Sweden, is the only device that utilizes 3D vector
sensing antennas and is readily able to detect these waves.  Other radio
telescopes, e.g., LOFAR \cite{LOFAR} and SKA \cite{SKA}, can be easily
reconfigured to achieve the same capability.

\section{Generalizing the results} \label{sec:cont}

From the previous Sections one might get the impression that the
analysis is specific to radio beams emitted by a circular antenna
array. However, as we now will show, the analysis is more general and
device independent.

Any device or source that generates a current distribution on a circle
\begin{align}
  \bfI(\rho,\Phi,z) = \frac{1}{2} \frac{N}{2\pi R} \frac{1}{\sqrt{2}}
  \begin{pmatrix} I \\ ihI \\ 0 \end{pmatrix} e^{il\Phi}
  \delta(\rho-R) \delta(z)\,,
\end{align}
where $\rho,\Phi$, and $z$ are cylindrical coordinates, results in the vector
potential
\begin{align}
  \bfA = \frac{1}{\sqrt{2}} \begin{pmatrix} 1 \\ ih \\ 0 \end{pmatrix}
  \calA(r) \calC(\theta,\phi)
\end{align}
with
\begin{align}
  \calC=\frac{N}{2\pi}\int d\Phi e^{il\Phi}
  e^{-ikR\sin\theta\cos(\Phi-\phi)}
\end{align}
and
\begin{align}
  \calA=-\frac{I\mu_0d}{8\pi}\frac{e^{ikr}}{r}\,,
\end{align}
cf.\ Eqs.~(\ref{A})--(\ref{C}). Hence, for the current distribution
given here, the results presented in the previous Sections are valid.

Since the angular dependence of the vector potential can be expressed by
its residue, Eq.~(\ref{C2}), there is no need for the current
distribution on the circle itself.  Mathematically speaking, the current
distribution can be along any path that encircles the origin, provided
that the residue is kept unchanged and that no further vortices
contribute.  Moreover, the current distribution can be chosen
arbitrarily in the $xy$ plane, provided that the residue is kept
unchanged.  Even if the residue changes, assuming that
Eqs.~(\ref{dCdphi}) and (\ref{dCdtheta}) remain valid, all results
presented in this paper, except Figs.~\ref{fig:2}--\ref{fig:5}, will
hold.

If the current distribution is not in a plane, but in three-dimensional
space, knotted optical vortices appear \cite{Flossmann&al:PRL:2008}.

\section{Interferences between waves associated with transverse and
  longitudinal currents}

If the vector potential reads
\begin{align}
  \bfA = \frac{1}{\sqrt{2}}
  \begin{pmatrix} \alpha \\ \alpha ih \\ \beta \end{pmatrix}
  \calA(r)\calC(\theta,\phi)\,,
\end{align}
the densities of energy and momenta are
\begin{gather}
  \wf=|\alpha|^2\wt+|\beta|^2\wl-2|\alpha\beta|\wi\,,
  \\
  \bfpf=|\alpha|^2\bfpt+|\beta|^2\bfpl-2|\alpha\beta|\bfpi\,,
  \\
  \bflf=|\alpha|^2\bflt+|\beta|^2\bfll-2|\alpha\beta|\bfli\,,
  \\
  \bfsf=|\alpha|^2\bfst+|\beta|^2\bfsl-2|\alpha\beta|\bfsi\,.
\end{gather}
The contributions associated with transverse currents are here
denoted with the upper index ``\transverse'' and are given in
Sections \ref{sec:w,p,j} and \ref{sec:s,l}, Eqs. (\ref{w})--(\ref{s}).
The contributions associated with longitudinal currents are given in
Section \ref{sec:long}, and the contributions due to interferences
read
\begin{gather}
  \wi = \frac{k^2}{4\mu_0} \Big(\frac{I\mu_0d}{8\pi r}\Big)^2
  \sin\theta\cos\theta (\calC^*\calC) \cos(h\phi+\arg\beta^*\alpha)\,,
\\
  \bfpi = \frac{\bfer k^2}{4c\mu_0} \Big(\frac{I\mu_0d}{8\pi r}\Big)^2
  \sin\theta\cos\theta (\calC^*\calC) \cos(h\phi+\arg\beta^*\alpha)\,,
\\
  \bfli = l \frac{-\bfetheta k^2}{4\omega\mu_0}
  \Big(\frac{I\mu_0d}{8\pi r}\Big)^2 \cos\theta (\calC^*\calC)
  \cos(h\phi+\arg\beta^*\alpha)\,, 
\end{gather}
and
\begin{multline}
  \bfsi = h \frac{-\bfetheta k^2}{4\omega\mu_0}
  \Big(\frac{I\mu_0d}{8\pi r}\Big)^2 \Big\{ -\cos\theta(\calC^*\calC)
  \\
  + \frac{1}{2}\sin\theta\frac{\partial}{\partial\theta}(\calC^*\calC)
  \Big\} \cos(h\phi+\arg\beta^*\alpha)
  \\
  + \frac{-\bfephi k^2}{4\omega\mu_0} \Big(\frac{I\mu_0d}{8\pi
    r}\Big)^2 (\calC^*\calC) \sin(h\phi+\arg\beta^*\alpha)\,,
\end{multline}
respectively.

If integrated, most of the interferences vanish. This allows to carry
over the principle of superposition of electromagnetic waves to their
mechanical properties,
\begin{gather}
  W = |\alpha|^2 \Wt + |\beta|^2 \Wl\,,
  \\
  \bfP = |\alpha|^2 \bfPt + |\beta|^2 \bfPl\,,
  \\
  \Lz = |\alpha|^2 \Lzt + |\beta|^2 \Lzl\,,
  \\
  \Sz = |\alpha|^2 \Szt, \quad \Szl = 0\,.
\end{gather}
However, for certain quantities the interferences do not cancel:
\begin{align}
  \Lxi \not= 0\,, \quad \Lyi \not= 0\,, \quad \Sxi \not= 0\,, \quad \Syi
  \not= 0\,.
\end{align}
The consequences of this is the subject of a forthcoming paper.

\section{Discrete sources and the uncertainty principle}

If, as in Section \ref{sec:cont}, the sources are continuous, the
analysis is exact in the far field and the number of orbital angular
momentum states is unbounded.
In contrast, if the sources are discrete, the presented
analysis is an approximation. Luckily, the approximation improves very
fast with the number of discrete sources, and there is typically no
need to worry about the accuracy.

However, for the case that only very few discrete sources are present,
or if one is interested in the number of orbital angular momentum states
that can be utilized, we propose an alternative approach.  Namely,
starting from a continuous source distribution the discrete nature of
the sources is imposed via a ``Pacman'' shaped mask that has slits at
the appropriate positions.  Due to Fourier convolution with the mask,
the uncertainty principle enters and gives rise to a distribution in the
orbital angular momentum
\cite{Franke-Arnold&al:NJP:2004,Rehacek&al:PRA:2008}.  This distribution
has a periodicity of $N$ and peaks of width proportional to $1/N$
limiting the number of orbital angular momentum states that can be
utilized in a circular array of $N$ equidistributed discrete sources to
\begin{align}
  \#\{l\} = N-O(1/N)\,,
\end{align}
where the implied constant in the last term depends on the
noise level and on the details of the discrete sources.

\section{Conclusions}

Our analysis of the properties of radio beams emitted by a circular
antenna array highlights analytically profound aspects of the beam spin
and orbital angular momentum.  Namely, how the beams are connected to
helicity and vorticity; that the angular momentum of the beams is an
intrinsic property; that it is transported all the way out to infinity;
that it can be separated into spin and orbital angular momentum; and
that spin, unlike its density, can be identified with polarization.

In particular, separating the angular momentum into spin and orbital
angular momentum, we find that the density of orbital angular momentum
depends on the vorticity only and can be expressed in quite simple
terms, whereas the density of spin angular momentum depends on both
helicity and vorticity.

The approach via an antenna array illuminates the physics from a new
perspective and confirms previous studies of orbital angular momentum of
electromagnetic beams
\cite{Barnett:JOB:2002,Allen&al:PO:1999,Allen&al:PRA:1992}.  Emerging as
a byproduct, but being of interest in its own, is the finding in Section
\ref{sec:long}.

\begin{acknowledgments}
Our sincerest thanks are due to Dr.~Tobia Carozzi for very valuable
comments and discussions.  Part of the work was supported by the Centre
for Dynamical Processes and Structure Formation, Uppsala University,
Sweden.  B.\,T.\ gratefully acknowledges financial support from the
Swedish Governmental Agency for Innovation Systems (VINNOVA) and the
Swedish Research Council (VR).
\end{acknowledgments}


\end{document}